\documentclass[aps,prb,twocolumn,showpacs,preprintnumbers,amsmath,amssymb,groupedaddress]{revtex4-1}
\usepackage{mathptm} 
\usepackage{dcolumn}                    
\usepackage{bm}                        
\usepackage{graphicx}
\usepackage{times}
\begin{document}
\newcommand {\be}{\begin{equation}}
\newcommand {\ee}{\end{equation}}
\newcommand {\bea}{\begin{eqnarray}}
\newcommand {\eea}{\end{eqnarray}}
\newcommand {\nn}{\nonumber}
\newcommand {\bb}{\bibitem}
\newcommand{\et}{{\,\it et al\,\,}}
\newcommand{\CuB}{CuBiS$_{2}$}
\newcommand{\BiTe}{Bi$_{2}$Te$_{3}$}

\title{Thermoelectric properties of AgGaTe$_2$ and related chalcopyrite structure materials}

\author{David Parker and David J. Singh}
\address{Oak Ridge National Laboratory, 1 Bethel Valley Rd., Oak Ridge, TN 37831}

\date{\today}

\begin{abstract}
We present an analysis of the potential thermoelectric performance of p-type AgGaTe$_{2}$, which has already shown a $ZT$ of 0.8 with partial optimization, and observe that the same band structure features, such as a mixture of light and heavy bands and isotropic transport, that lead to this good performance are present in certain other ternary chalcopyrite structure semiconductors. We find that optimal performance of AgGaTe$_2$ will be found for hole concentrations between 4 $\times 10^{19}$ and 2 $\times 10^{20}$cm$^{-3}$ at 900 K, and 2 $\times 10^{19}$ and 10$^{20}$ cm$^{-3}$ at 700 K, and that certain other chalcopyrite semiconductors might show good thermoelectric performance at similar doping ranges and temperatures if not for higher lattice thermal conductivity.

\end{abstract}
\maketitle

\section{Introduction} 

Thermoelectric performance, as characterized by the so-called figure of merit,
$ZT$, is a property of matter
that has attracted much recent interest.
The expression for $ZT$ is $ZT=\sigma S^2 T/\kappa$, where $S$ is the thermopower,
$T$ is temperature, $\sigma$ is electrical conductivity and
$\kappa$ is the thermal conductivity, typically written as the sum
of lattice and electronic contributions, $\kappa=\kappa_l+\kappa_e$.
Obtaining high $ZT$ is a fundamental scientific challenge, since
high $ZT$ is a strongly counter-indicated transport property.
Specifically, one requires (1) high mobility at the same time as low thermal
conductivity, suggesting weak scattering of charge carriers, but strong
scattering of phonons, (2) high conductivity and high thermopower, (3)
low thermal conductivity (i.e. soft lattice) and high melting point, and
finally (4) the combination of heavy band behavior (for high $S$) at the same
time as effective controlled doping.
Although there is no known fundamental limit on $ZT$, for many decades
the maximum known $ZT$ in any material was near 1.0, while in recent
years new concepts such as the use
of nanostructuring \cite{poudel,tritt}or `rattling' \cite{sales,subramian,shi} to reduce thermal conductivity,
and complex or modified electronic structure (e.g. by nanoscale effects \cite{hicks},
or selection of materials with unusual
band structures \cite{sny_amat, subramian, hsu}) have raised the best $ZT$ values to near 2.  Reviews of the field may be found in
Refs. \onlinecite{sny_tob,dress}.

Here we discuss AgGaTe$_2$ and related chalcopyrite compounds, which
we find to have unusual band structures combining heavy and light
features that represent one route for resolving the above conundrums,
particularly those relating to electrical conductivity and thermopower.  An early study of chalcopyrite band structures is found in Ref. \onlinecite{zunger}.

While heavy mass bands are generally favorable towards producing high thermopower, an essential ingredient for thermoelectric performance, such bands also generally reduce the carrier mobility and conductivity, so that very heavy mass bands on their own are not universally beneficial for good thermoelectric performance (i.e. {\it ZT}).  Light mass bands, by contrast, are favorable for electrical conductivity but not so for thermopower. However, a {\it mixture} of light and heavy bands has previously been shown to be beneficial for thermoelectric performance \cite{mazin_singh}, with the light band providing good conduction and the heavy band a small energy scale helpful for the thermopower.  The telluride La$_{3}$Te$_{4}$ is a good example \cite{may} of a high performance material in which this behavior is realized.


AgGaTe$_{2}$ has already shown a $ZT$ of 0.8 at 850 K \cite{yusufu} at a low hole doping of approximately 10$^{16}$cm$^{-3}$, far outside the heavy doping ranges of 10$^{19}$-10$^{21}$cm$^{-3}$ where optimal performance is typically found in thermoelectrics.  Structurally, it is very different from the chemically related compounds, PbTe and AgSbTe$_2$, as it is tetrahedrally bonded, rather than octahedral. We show that the valence band electronic structure of this $p$-type material is very similar to that of certain other ternary chalcopyrite semiconductors, which also show two factors favorable for thermoelectric performance - nearly isotropic transport and a mixture of heavy and light bands. The nearly isotropic  transport arises from the (well known) similarity of the chalcopyrite physical structure of these materials and the cubic zincblende structure, as described above.  We depict the chalcopyrite structure of AgGaTe$_{2}$ in Figure 1.

\begin{figure}[h!]
\includegraphics[width=8cm]{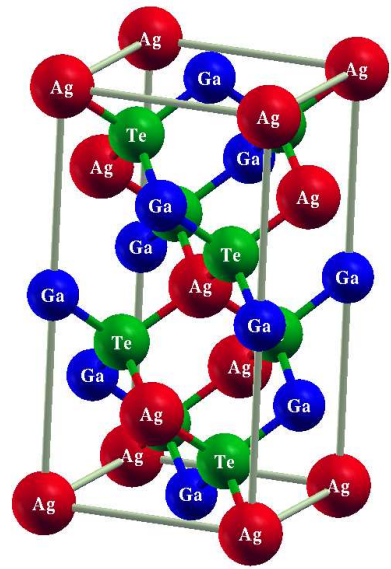}
\caption{The physical structure of AgGaTe$_2$.  The planar lattice constant is 6.23 \AA\, and the $c$-axis value 11.96 \AA\, for a c/a ratio of 1.92.}
\end{figure}

\section{Model, calculated bandstructures and density-of-states}

In order to study the transport in AgGaTe$_2$ and related materials quantitatively, we employ first principles density-functional theory as implemented in the linearized augmented plane-wave (LAPW) WIEN2K code \cite{wien}.   Since first principles methods often significantly understate the band gap, we here employ a modification
of the generalized gradient approximation (GGA) due to Tran and Blaha \cite{tran} known as a modified Becke Johnson potential \cite{becke}, which has been shown to give much more accurate band gaps than the standard GGA.  
All calculations were performed to self-consistency to an accuracy of better than than one meV per unit cell, using between 1000 and 2000 $k$-points in the full Brillouin zone, with spin-orbit coupling included for all materials except for ZnSiAs$_2$.  For AgGaTe$_{2}$, internal coordinates were optimized, using the Perdew-Burke-Ernzerhof \cite{perdew} GGA.  To calculate the thermopower, as well as the conductivity anisotropy, we used the Boltzmann transport code BOLTZTRAP \cite{madsen} within the constant scattering time approximation (CSTA).  In this approximation, the scattering time of an electron is assumed to depend only on doping and temperature, but not the {\it energy} of the electron.  When employed within the canonical expressions for the thermopower and conductivity, as given in 
Ref. \onlinecite{ziman}, this results in expressions - the thermopower S and the conductivity anisotropy $\sigma_{planar}/\sigma_{c-axis}$ - with {\it no} dependence on any assumed absolute value of
the scattering time.  The CSTA has been used with quantitative success to describe the thermopower of a large number of thermoelectric materials \cite{zhang3,parker,singh_PbTe,ong,madsen2,singh3,scheide, bertini,lykke,wang}.

We begin with the band structure, previously considered in Ref. \onlinecite{reshak}.   Depicted in Figure 2 is the calculated bandstructure of AgGaTe$_{2}$ within the tetragonal Brillouin zone \cite{lax} .  The calculated band gap, at 1.15 eV, falls in the center of the 0.9 - 1.3 eV range of band gap values found in the literature \cite{tell,kumar,roy,schunemann,chatraphorn,arai,pamplin,shewchun}, and is sufficient  to prevent bipolar conduction at temperatures of 900 K and below.

Both the valence band maximum (VBM)
and conduction band minimum (CBM) are located at the $\Gamma$ point.  These pockets generally show a fair degree of isotropy, with the dispersion somewhat greater along
the $\Gamma$-Z line than the planar $\Gamma$-N direction.  Of interest for the thermoelectric performance, the plot depicts a mixture of heavy and light bands near the VBM.  
The heavy band shows a $\Gamma$-N dispersion of 0.7 eV, leading to an approximate band mass of 1 $m_{0}$, with the light band at roughly half this mass.  As stated previously, this light-band/heavy band combination
has previously been shown to be good for thermoelectric performance \cite{mazin_singh}.

\begin{figure}[h!]
\includegraphics[width=8cm]{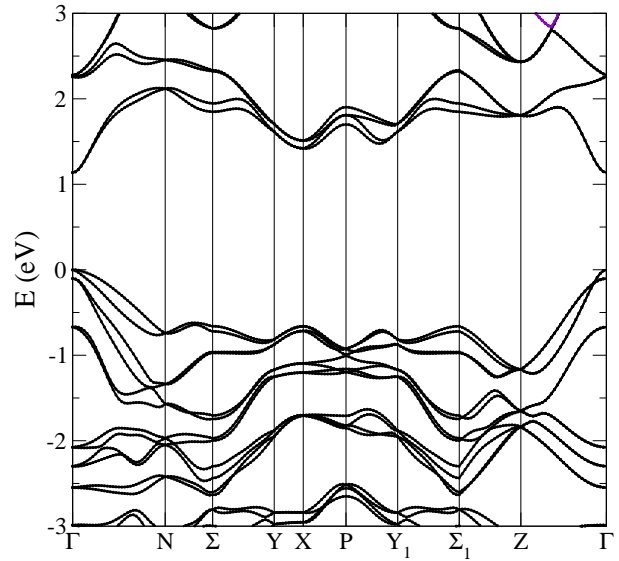}
\caption{The bandstructure of AgGaTe$_{2}$.  The zero of energy is set to the valence band maximum.}%
\end{figure}
In Figure 3 we present the calculated density-of-states.  The heavy valence band's impact is immediately apparent, with the DOS rising rapidly just below the VBM.  A similarly heavy band appears somewhat above the 
CBM, with a highly dispersive band (inset) in the first 0.25 eV above the CBM.
Also presented in Figure 3 is  the atom-projected DOS. It is worth noting that for all atoms and both the VBM and CBM, the relevant atomic character is of virtually the same shape as the overall DOS, suggesting a coherence to the electronic scattering which tends to affirm the accuracy of the CSTA.

\begin{figure}[h!]
\includegraphics[width=8cm]{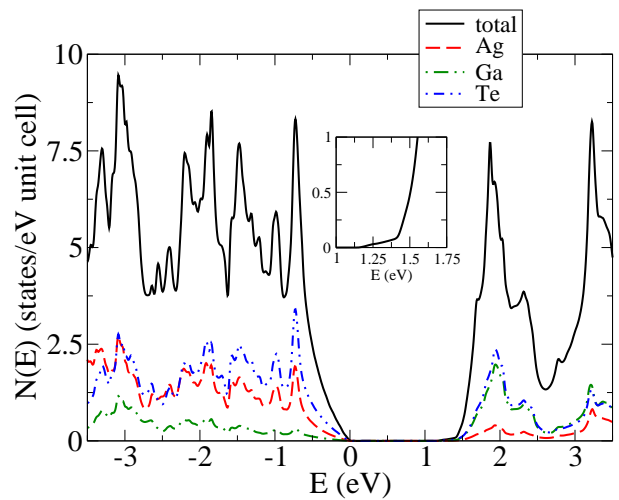}
\caption{The density of states of AgGaTe$_{2}$.  The zero of energy is set to the valence band maximum. Inset: the density of states near the conduction band minimum.} 
\end{figure}

\section{Calculated thermopower and conductivity results}

In Figure 4 we depict the calculated hole-doped thermopower results for AgGaTe$_{2}$.  The plot depicts (at 700 and 900 K, the maximum operation temperature for AgGaTe$_{2}$) an essentially logarithmic dependence of thermopower on carrier concentration, in line with Pisarenko  behavior.  No effects of bipolar conduction are visible, and the plot shows 900 K thermpowers (virtually isotropic as described in Section 1) approaching 400 ${\mathrm \mu}$V/K at hole concentrations p of $2 \times 10^{19}$cm$^{-3}$.  Given the lack of information regarding the hole mobility at these concentrations and temperatures, estimating the figure-of-merit $ZT$ at these temperatures is impractical.  We can say, however, that in previously studied materials thermoelectric performance is typically optimum for thermopowers between 200-300 ${\mathrm \mu}$V/K.  Note that the Wiedemann-Franz relationship implies that, even if the lattice thermal conductivity were nil, a thermopower of 156${\mathrm \mu}$V/K
would be required to attain a $ZT$ of unity (the typical minimum value for a material to be considered a ``high performance thermoelectric"), so that in practice thermopowers substantially above this value are necessary to achieve high performance.  

At 900 K for AgGaTe$_{2}$ these 200-300 ${\mathrm \mu}$V/K thermopowers are found for hole concentrations between 4 $\times 10^{19}$ and 2 $\times 10^{20}$cm$^{-3}$; at 700 K these thermopowers are found for concentrations between 2 $\times 10^{19}$ and 10$^{20}$ cm$^{-3}$.  While we cannot make a definite estimate of $ZT$, we can say with high confidence that performance substantially above the 0.8 $ZT$ value achieved in Ref. \onlinecite{yusufu} will be found.  We assert this because the sample in this reference was sufficiently underdoped as to yield a thermopower which {\it decreased} with increasing temperature from 300 K all the way to the highest temperature measured, strongly indicative of bipolar conduction (always unfavorable for thermoelectric performance).  Our results here show bipolar conduction can be avoided by heavy doping while maintaining high thermopower.
\begin{figure}[h!]
\includegraphics[width=8cm]{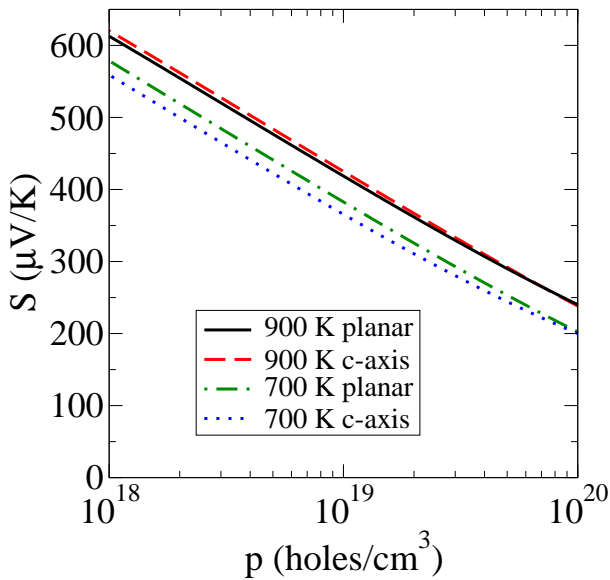}
\caption{The calculated hole-doped thermopowers of AgGaTe$_{2}$ at 700 and 900 K.}
\end{figure}

Although experimental work to date has found that AgGaTe$_{2}$ tends to form $p$-type, in Figure 5 we present the calculated electron doped thermopower.  Somewhat lower values than in the hole-doped case are apparent due to the increased dispersion on the electron-doped side, but the values depicted are still substantial.  In addition, as with the valence bands the conduction bands contain a mixture of heavy and light bands, beneficial for thermoelectric performance.  Finally, although the thermopower is lower than for hole doping, this can be partly compensated for by the likely increased mobility for electron doping.  We therefore expect that good performance may obtain for electron doping, in the range of $ 1.5 \times 10^{19}$  - 10$^{20}$ cm$^{-3}$ at 900 K and $3 \times 10^{18}$  - 2$\times 10^{19}$ cm$^{-3}$ at 700 K (using the same criteria as for hole doping).
\begin{figure}[h!]
\includegraphics[width=8cm]{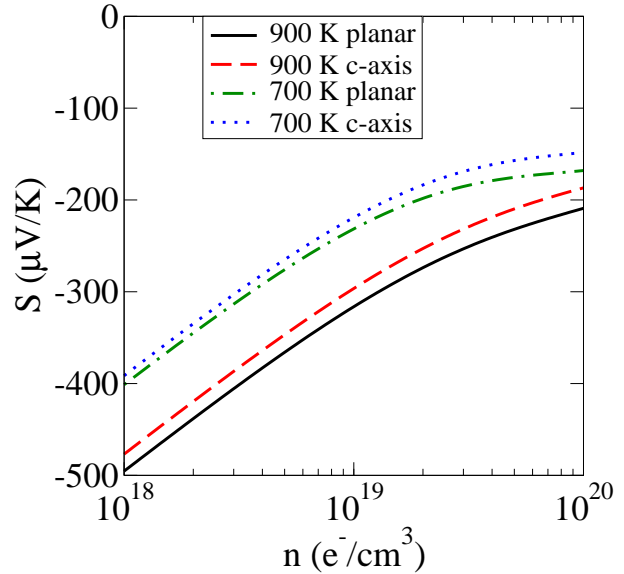}
\caption{The calculated electron-doped thermopowers of AgGaTe$_{2}$ at 700 and 900 K.}
\end{figure}

In Figure 6 we depict the hole-doped thermopower at 300 K.  This shows similarly favorable behavior to the high-temperature results, albeit with lower values.  Somewhat greater anisotropy than in the high temperature case is apparent, due mainly to the narrower energy range of the valence band that is relevant for transport at these lower temperatures.  This can be seen more directly from looking at the band structure plot (Fig. 2) - the band mass of the light band in the $\Gamma$-Z direction is roughly one half the mass of the heavy band in the $\Gamma$-N direction.  At low dopings and temperatures (such as 300 K), for c-axis transport it is only this light band that is operative in transport and there is therefore substantial anisotropy in the calculated thermopower. As one moves to heavier dopings the heavier band (whose maximum is roughly 100 meV below the VBM) becomes operative and yields substantially more isotropic behavior.  We note also that both planar and c-axis thermopower obey a Pisarenko type relationship (i.e. logarithmic in carrier concentration) at the low dopings, indicative of non-degenerate single band transport, and deviate from this at higher dopings, due both to the two band behavior and the approaching of the degenerate limit. 
\begin{figure}[h!]
\includegraphics[width=8cm]{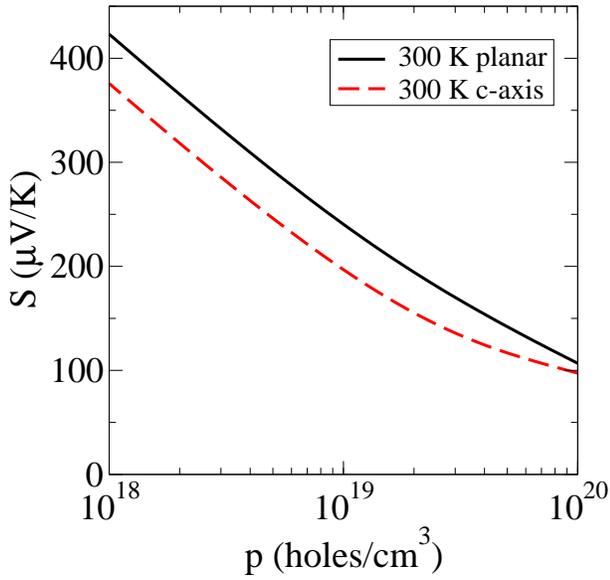}
\caption{The calculated hole-doped thermopowers of AgGaTe$_{2}$ at 300 K.}
\end{figure}
In Figure 7 we present the 900 K conductivity anisotropy, which is essentially nil, a significant advantage for applications, as discussed previously.
\begin{figure}[h!]
\includegraphics[width=8cm]{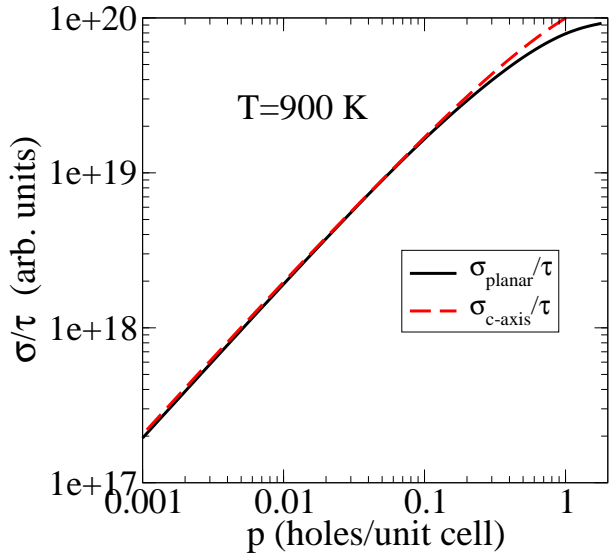}
\caption{The calculated electrical conductivity (divided by scattering time) of AgGaTe$_{2}$ at 900 K.  Note that one hole per unit cell is equivalent to 4.31 $\times 10^{21}$ cm$^{-3}$.}
\end{figure}


\section{Observation on Valence Band Structure in Ternary Chalcopyrite Semiconductors and Thermopower of CdGeAs$_2$}

In this section we point out that there are a number of ternary chalcopyrite semiconductors with nearly the same physical and electronic structure as AgGaTe$_{2}$ that can be expected to give similarly beneficial Seebeck coefficients and isotropic electronic conductivity.  To illustrate the point in Figure 8 we present the calculated band structure of {\it six} different chalcopyrite structure semiconductors, including AgGaTe$_{2}$.  For simplicity we limit ourselves to the valence band structure as most, if not all these compounds generally behave as p-type semiconductors.  We note firstly that in all these compounds the valence band maximum is centered at $\Gamma$ (note that the plots are scaled so that within a given plot, momentum space distances between labeled points are proportional to the distance along the x-axis).  The plots also indicate a general consistency of dispersive energy scales - for all of the plots the $\Gamma$-N dispersion is between 0.4 and 0.8 eV and that in the $\Gamma$-Z direction betwen 0.75 eV and 1.2 eV.  While the plot-to-plot differences increase at greater distances ($>$ 1 eV)  from the VBM, for the purposes of transport consideration these are of little importance.
\begin{figure}[h!]
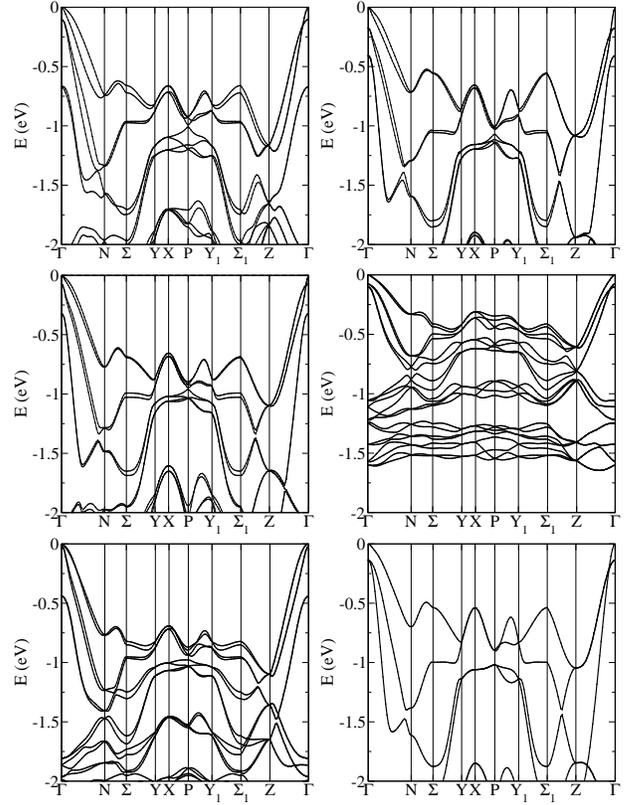

\includegraphics[width=4cm]{Fig8a.eps}
\includegraphics[width=4cm]{Fig8b.eps}
\includegraphics[width=4cm]{Fig8c.eps}
\includegraphics[width=4cm]{Fig8d.eps}
\includegraphics[width=4cm]{Fig8e.eps}
\includegraphics[width=4cm]{Fig8f.eps}
\caption{The calculated valence band structures of (top left) AgGaTe$_{2}$; top right CdGeAs$_{2}$; middle left CdSnAs$_{2}$; middle right CuInS$_2$; bottom left CuInTe$_{2}$; bottom right ZnSiAs$_{2}$. }\end{figure}

As with AgGaTe$_2$, these valence band plots generally contain a mixture of heavy and light bands.  This has previously been shown \cite{mazin_singh, may} to be good for thermoelectric performance.  However,  the lattice thermal conductivity (presented in Table 1) of most of these materials is much higher than that of AgGaTe$_{2}$, making them generally less favorable for thermoelectric performance.  To the degree that this lattice term can be reduced by alloying and nanostructuring, these materials may show good performance as well.   Given the large number of materials with very similar electronic structure, we would expect that alloying these materials with each other should be possible and effective at reducing $\kappa_{lattice}$.  We note (Table 1)  that for these materials, with the exception of  CdSnAs$_2$, the experimental band gap is sufficiently large to prevent bipolar conduction at temperatures below 900 K, so that the assessment of favorable valence band structure implies good thermopower behavior as well.

Perhaps the likely best performer of the remaining five materials  would be CdGeAs$_2$ with its 300 K lattice thermal conductivity listed in Ref. \onlinecite{spitzer} as 4 W/m-K.  In Figure 9 we present the calculated 900 K hole-doped thermopower (the melting point is 943 K) of this material, noting that even at the relatively high hole doping of 10$^{20}$cm$^{-3}$ the thermopower is still over 250 ${\mathrm \mu}$ V /K, an excellent value for thermoelectric performance, particularly since the lattice term at this temperature (assuming a canonical $1/T$ behavior) would be just 1.3 W/m-K. 
\begin{table}[tbp]
\caption{Lattice thermal conductivity $\kappa_{lattice}$ at 300 K and experimental band gaps of chalcopyrite compounds. Thermal conductivity (taken for polycrystalline samples) from Ref. \onlinecite{spitzer} , and band gaps from Ref. \onlinecite{springer}, except where noted.}
\begin{center}
\begin{tabular}{|c|c|c|}
\hline
Compound  &  $\kappa_{lattice}$  (W/m  K) & Band gap (eV)\\ \hline
AgGaTe$_2$  &  1.7 \cite{yusufu} & 1.15\\ \hline
CdGeAs$_{2}$ & 4.0 & 0.57 \\ \hline
CdSnAs$_{2}$ & 7.5 & 0.26 \cite{footnote} \\ \hline
CuInS$_{2}$ & 12.5\cite{mak} & 1.53 \\ \hline
CuInTe$_{2}$ & 5.5 & 0.98 \\ \hline
ZnSiAs$_{2}$ & 14.0 & 1.92 \\ \hline

\end{tabular}
\end{center}
\end{table}

 Included in the plot are two calculated curves - one assuming the first principles calculated band gap of 0.65 eV, and a results assuming a somewhat smaller gap of 0.5 eV.  We have included the additional curve because there is evidence \cite{springer} that the band gap of CdGeAs$_2$ decreases significantly with temperature.  For both curves, as with the AgGaTe$_{2}$ the concentration dependence is essentially logarithmic at high dopings, until one approaches the bipolar regime where the thermopower decreases with decreasing concentration.  For the as-calculated gap of 0.65 eV this happens for $p=3 \times
10^{18}$cm$^{-3}$; for the smaller gap this happens at $10^{19}$cm$^{-3}$. Using the same criteria as for AgGaTe$_2$ we find that optimal doping will most likely be found for hole concentrations between $5 \times 10^{19}$ and $2 \times 10^{20}$cm$^{-3}$, and this statement is independent of the value taken for the band gap.  We note also that in the non-bipolar regime (i.e. to the right of the thermopower maximum in Figure 9) the thermopower is very similar to that of AgGaTe$_2$, as would be expected given the similar electronic structure.
\begin{figure}[h!]
\vspace{1cm}
\includegraphics[width=8cm]{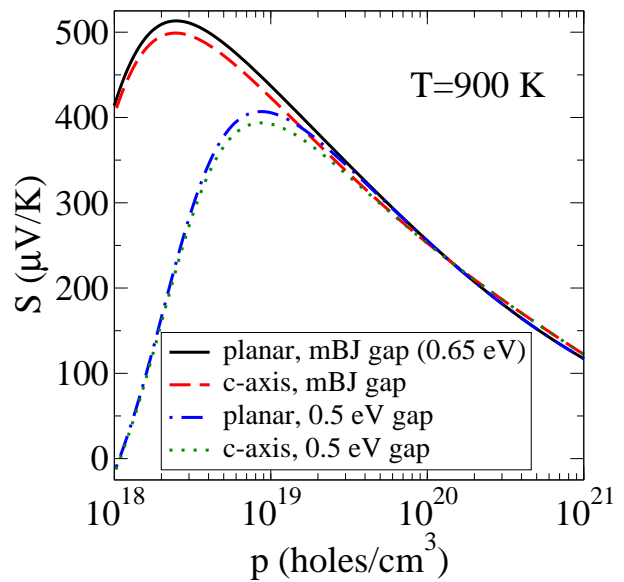}
\caption{The calculated hole-doped thermopower of CdGeAs$_{2}$ at 900 K.}
\end{figure}

Although we have not calculated the thermopower of the remaining materials, to the extent that the dispersive energy scales are similar to those of AgGaTe$_2$ and CdGeAs$_2$ the thermopower will be similar as well. CuInS$_2$, in particular, may well have even larger thermopower than these materials given the smaller $\Gamma$ -N and $\Gamma$-Z dispersions; actual performance of this material, however, is expected to be low due to the high lattice thermal conductivity and likely low mobility of this sulfide; the same considerations apply to ZnSiAs$_2$.

\section{Conclusion}

To conclude, in this work we have shown that the ternary chalcopyrite semiconductor AgGaTe$_2$ may show excellent thermoelectric performance at hole dopings ranging from 4 $\times 10^{19}$ and 2 $\times 10^{20}$cm$^{-3}$ at 900 K and between 2 $\times 10^{19}$ and 10$^{20}$ cm$^{-3}$ at 700 K.   This performance may well be due to a heavy-band light-band structure near the valence band maximum and will be aided by nearly isotropic transport.  In addition, we have shown that the valence band structure of this material is very similar to that of a number of ternary chalcopyrite semiconductors, which might therefore show good thermoelectric performance if not for a relatively high lattice thermal conductivity.  Given the general alloying capability of chalcopyrite semiconductors, it may be of interest to pursue heavy doping of these materials in concert with alloying with other chalcopyrite materials.
\\
\\
{\bf Acknowledgments} 

DJS is grateful for helpful discussions on tetrahedrally bonded thermoelectric materials with X. Shi, Lili Xi, Jiong Yang and Wenqing Zhang.  This research was supported by the U.S. Department of Energy, EERE, Vehicle Technologies, Propulsion Materials Program (DP) and the ÔSolid State Solar-Thermal Energy Conversion Center (S3 TEC)Õ, an Energy Frontier Research Center funded by the US Department of Energy, Office of Science, Office of Basic Energy Sciences under Award Number: DE-SC0001299/DE-FG02-09ER46577 (DJS).

\end{document}